\documentclass[preprints,article,submit,pdftex,moreauthors]{Definitions/mdpi} 

\usepackage{cite}
\usepackage{amsmath,amssymb,amsfonts}
\usepackage{algorithmic}
\usepackage{graphicx,color}
\usepackage{graphicx}
\usepackage{textcomp}

\usepackage[nolist]{acronym}
\usepackage[colorinlistoftodos, textsize=footnotesize, textwidth=20mm]{todonotes}
\usepackage{svg} 
\usepackage{siunitx} 

\newcolumntype{M}[1]{>{\centering\arraybackslash}m{#1}} 
\newcolumntype{P}[1]{>{\raggedright\arraybackslash}p{#1}} 

\newcommand{\tsuper}{\textsuperscript}
\newcommand{\tsub}{\textsubscript}

\newcommand{\guis}{GUI\textsubscript{Static}\,}
\newcommand{\guid}{GUI\textsubscript{Dynamic}\,}

\firstpage{1} 
\pubvolume{1}
\issuenum{1}
\articlenumber{0}
\pubyear{2025}
\copyrightyear{2025}
\datereceived{ } 
\daterevised{ } 
\dateaccepted{ } 
\datepublished{ } 
\hreflink{https://doi.org/} 

\Title{A User-Centered Teleoperation GUI for Automated Vehicles: 
Identifying and Evaluating Information Requirements for Remote Driving and Assistance}

\TitleCitation{A User-Centered Teleoperation GUI for Automated Vehicles: Identifying and Evaluating Information Requirements for Remote Driving and Assistance}

\Author{Maria-Magdalena Wolf *\orcidA{}, Henrik Schmidt\orcidB{}, Michael Christl\orcidC{}, Jana Fank\orcidD{} and Frank Diermeyer\orcidE{}}

\AuthorNames{Maria-Magdalena Wolf, Henrik Schmidt, Michael Christl, Jana Fank and Frank Diermeyer}

\isAPAStyle{%
       \AuthorCitation{Wolf, M.-M., Schmidt, H., Christl, M., Fank J. \& Diermeyer, F.}
         }{%
        \isChicagoStyle{%
        \AuthorCitation{Wolf, Maria-Magdalena, Henrik Schmidt, Michael Christl, Jana Fank, and Frank Diermeyer.}
        }{
        \AuthorCitation{Wolf, M.-M.; Schmidt, H.; Christl, M.; Fank J.; Diermeyer, F.}
        }
}

\address[1]{%
Institute of Automotive Technology, Technical University of Munich (TUM), \\ 
Boltzmannstr. 15, DE-85748 Garching b. M\"unchen, Germany} 

\corres{Correspondence: maria.wolf@tum.de}

\abstract{
Teleoperation emerged as a promising fallback for situations beyond the capabilities of automated vehicles. 
Nevertheless, teleoperation still faces challenges, such as reduced situational awareness. 
Since situational awareness is primarily built through the remote operator’s visual perception, the Graphical User Interface (GUI) design is critical. 
In addition to video feeds, supplemental informational elements are crucial - not only for the predominantly studied Remote Driving 
but also for the arising desk-based Remote Assistance concepts. 
This work develops a GUI for different teleoperation concepts by identifying key informational elements during the teleoperation process through expert interviews (N~=~9). 
Following this, a static and dynamic GUI prototype is developed and evaluated in a click-dummy study (N~=~36). 
Thereby, the dynamic GUI adapts the number of displayed elements according to the teleoperation phase.
Results show that both GUIs achieve good System Usability Scale (SUS) ratings, with the dynamic GUI significantly outperforming the static version in both usability and task completion time. 
The User Experience Questionnaire (UEQ) score shows potential for improvement. 
To enhance the user experience, the GUI should be evaluated in a follow-up study that includes interaction with a real vehicle.}

\begin{document}

\begin{acronym}

    \acro{ATI}[ATI]{Affinity for Technology Interaction}

    \acro{av}[AV]{Automated Vehicle}
    \acroplural{av}[AVs]{Automated Vehicles}

    \acro{gui}[GUI]{Graphical User Interface}
    \acroplural{gui}[GUIs]{Graphical User Interfaces}


    \acro{sus}[SUS]{System Usability Scale}

    \acro{ueq}[UEQ]{User Experience Questionnaire}

\end{acronym}



\section{Introduction}
\label{introduction}


As of 2025, driverless cars remain a rare sight, especially in Europe, even though they exist and are permitted by legislation \citep{eu-1426-2022}. 
However, if there is an \ac{av} on public roads, chances are high that a human is still in the loop. 
Major \ac{av} companies like Waymo \citep{Waymo2025} or Cruise \citep{Cruise2025} continue to rely on remote operators for monitoring, guidance, and driving \citep{Jin2021}. 
From a car user perspective, attitudes toward autonomous and remote driving also vary. 
A McKinsey survey \citep{Kelkar2025} of around 1,500 car owners in China, Germany, and the United States found that consumers consider remote driving superior to autonomous driving in several aspects, including safety and adaptability to unexpected situations. 
Additionally, about 56\,{\%} of respondents indicated they would more likely purchase an \ac{av} if it included remote-driving features, as this could allow for broader usability across different road types and weather conditions. 
Thus, besides automation technology companies, actual teleoperation providers like Valeo \citep{Valeo2023}, DriveU \citep{DriveU2025}, Einride \citep{Einride2024} or Fernride \citep{Fernride2024} appeared. 

If human involvement is essential for autonomous driving, so is the user interface for interacting with the vehicle. 
In addition to intuitive interaction, situational awareness of the remote operator presents a particular challenge in teleoperation \citep{Mutzenich2021, Carsten2020}. 
Since situational awareness relies mainly on the remote operator's visual perception \citep{Colavita1974}, particular emphasis is placed on designing the \ac{gui} to effectively convey essential information to the remote operator, enabling quick and safe decision-making and acting.

Therefore, our work aims to develop a user-friendly \ac{gui} focusing on necessary informational elements complementing the video feed for safe operations by following the user-centered design process by: 
\begin{itemize}
    \item Defining teleoperation process steps for Remote Driving and Remote Assistance.
    \item Identifying essential \ac{gui} elements through expert interviews. 
    \item Evaluating a static and dynamic variant of the developed \ac{gui} within the teleoperation process through an online study.
\end{itemize}

Typically, teleoperation \acp{gui} present a camera feed along with additional information like the vehicle speed. 
Several studies have already explored optimal video presentation in teleoperation, focusing on aspects such as required video quality \citep{Geo2020b}, camera perspective~\citep{Boker2023}, and field of view \citep{Voysys2020}. 
In addition, research has examined a range of display concepts \citep{Geo2020b, Cab2019} and output devices, including head-mounted displays \citep{Bou2017, Georg2024}.

In contrast, this work focuses
on the supplemental informational elements and how a holistic \ac{gui} design should look on a conventional monitor to accommodate all necessary information on a single display surface. 
The following summarizes the literature on evaluating teleoperation \acp{gui} and deriving design recommendations.
Table~\ref{table:related_work} provides an overview of the teleoperation concepts considered in the publications, the implementation approaches, setups (including output devices, displayed elements, and input devices), and participants. 
The table also distinguishes whether the evaluations focused on the overall \ac{gui} or individual display elements.

\begin{table*}[htbp]
\caption{Overview of Related Work regarding Graphical User Interfaces for Teleoperation}
\footnotesize
\renewcommand{\arraystretch}{1.2} 
\begin{center}
\begin{tabular}{|P{0.15cm} |M{1.0cm}|M{0.9cm}|M{0.9cm}|M{0.9cm}|M{1.1cm}|M{2.2cm}|M{1.2cm}|M{0.4cm}|M{1.6cm}|M{1.2cm}|M{1.0cm}|}
\cline{2-12}
\multicolumn{1}{c|}{}& \multirow{2}{0.9cm}{\centering \rule{0pt}{16pt}\textbf{Source}} & \multicolumn{3}{c|}{\textbf{Implementation}} & \multicolumn{3}{c|}{\textbf{Setup}} & \multicolumn{2}{c|}{\textbf{Participants}} & \multicolumn{2}{c|}{\textbf{Evaluation}} \\ 
\cline{3-12}
\multicolumn{1}{c|}{} & & \textbf{None} & \textbf{Simu-lation} & \textbf{Real World}&\textbf{Output Device}&\textbf{Display Elements}&\textbf{Input Device}&\textbf{No.}&\textbf{Experience}&\textbf{Display Elements}&\textbf{Overall GUI}\\
\hline
\multirow{5}{*}{\rotatebox{90}{\parbox{12cm}{\centering \textbf{Remote Driving}}}} & Tener and Lanir \citep{Tener2022} & Observa-tions/ Interviews & & & Ultrawide Monitor & video feeds, side mirrors, rear view, vehicle speed, other driving information $\rightarrow$ derive 25 teleoperation challenges and design guidelines & Steering wheel and pedals & 8/ 14 & Remote Operators/ Automotive industry professionals, academic teleoperation researchers & (\checkmark) & \\
\cline{2-12}
&Graf and Hussmann \citep{Graf2020} & Inter- views & & & & 80 collected requirements/
20 extracted for safe AV teleoperation & & 18/ 10 & Automotive industry professionals & \checkmark & \\
\cline{2-12}
& Gafert et al. \citep{Gafert2023} & & &Mini-ature Vehicle& Head-Mounted
Display& 36 requirements represented
in 19 GUI elements:
video feeds, rear view, navigation map, vehicle speed, other driving information, task information, weather, … & Steering wheel and pedals & 16 & None & \checkmark & \checkmark \\
\cline{2-12}
& Bodell and Gulliksson \citep{Bodell2016} & &Gazebo& &2 Monitors&360° video feed,  vehicle speed,
bird's-eye view map, planned path of automation, LiDAR data&Steering wheel, Gamepad controller&?&None&\checkmark&\checkmark \\
\cline{2-12}
& Lindgren and Vahlberg \citep{lindgren2023} & &Click-Dummy & &3 Monitors&video feeds, side mirrors, rear view, bird's eye view, vehicle speed, other driving information, notification list,  top bar with system information, time and weather
$\rightarrow$ derive general design guidelines&Mouse&9 &Engineers with experience in remote operation,
Remote Operators& &\checkmark\\
\hline
\multirow{3}{*}{\rotatebox{90}{\parbox{6cm}{\centering \textbf{Remote Assistance}}}}& Kettwich et al. \citep{Kettwich2021} & &Click-Dummy & &6 Monitors, Touchscreen&video feeds, overview map, vehicle data, disturbances; bird's eye view, planned path of automation&Mouse, Touch&13&Control center professionals& &\checkmark \\
\cline{2-12}
& Schrank et al. \citep{Schrank2024} & &Click-Dummy & &6 Monitors, Touchscreen&video feeds, overview map, vehicle data, disturbances; bird's eye view, planned path of automation&Mouse, Touch&34&University or state-certified technical degree (remote operator criteria acc. to German law)& &\checkmark \\
\cline{2-12}
& Tener and Lanir \citep{Tener2025} & &Click-Dummy& &Ultrawide Monitor,
Monitor on top, Tablet&video feeds; rear view;
video feed, status bar, planned path of automation, obstacle marking, notifications, vehicle status, weather, …&Touch&14&Experts in remote operation& &\checkmark\\
\hline
\end{tabular}
\label{table:related_work}
\end{center}
\end{table*}

Based on expert interviews or observations, publications propose design recommendations or requirements for teleoperation interfaces without presenting a fully developed \ac{gui} in their work \citep{Graf2020, Tener2022}. 
For instance, Graf and Hussmann \citep{Graf2020} identified 80 general information requirements, 
such as object display and vehicle position, which they reduced to 20 essential elements deemed critical for safe teleoperation. 
In contrast, Tener and Lanir \citep{Tener2022} illustrate their \ac{gui} element suggestions in design examples. However, the proposed informational elements are neither integrated into a unified \ac{gui} nor empirically validated. 

While the work by Lindgren and Vahlberg \citep{lindgren2023} presents a holistic \ac{gui}, their study evaluates a click-dummy and derives general design guidelines, including information prioritization and the use of coding strategies like color to enhance clarity. 
Nevertheless, their work does not explicitly examine the impact of individual \ac{gui} elements. 

Further developing this line of research, Gafert et al. \citep{Gafert2023} and Bodell and Gulliksson~\citep{Bodell2016} shift the focus toward analyzing individual \ac{gui} components. 
Gafert et al. \citep{Gafert2023} are the first to translate the requirements defined by Graf and Hussmann \citep{Graf2020} into 19 concrete \ac{gui} elements, which are then implemented in two display variants: one featuring the complete set of informational elements and the other encompassing a reduced display configuration. 
In a real driving study involving the teleoperation of a miniature vehicle, the density and relevance of displayed information are examined across the orientation and navigation phases. 
Bodell and Gulliksson \citep{Bodell2016} also conducted a user study to explore how camera images, maps, and sensor data can be effectively presented to enhance safety and efficiency. 
Other informational display elements, such as steering angle, turn signal indicators, or vehicle model information, were not included in their \ac{gui} and evaluation.

The publications presented so far all focus on Remote Driving. 
In addition to traditional Remote Driving workstations, startups and research initiatives are now introducing novel teleoperation workstations with alternative desk-based control and display concepts, primarily for Remote Assistance of the \ac{av}. 
Click-based Remote Assistance systems offer an intuitive way to issue high-level commands, with users favoring minimal involvement in the dynamic driving task unless the 
scenario feels unsafe \citep{Brand2024, Brecht2024}. 
This requires different user interface specifications \citep{Schrank2024a} and 
research in Remote Assistance \citep{Skogsmo2023}.

Implementations by Cruise~\citep{Cruise2021}, Motional~\citep{Motional2022, Motional2023}, and Zoox~\citep{Zoox2020} 
display an abstract representation of the surroundings next to multiple video streams, including road layout and detected objects. Motional~\citep{Motional2022} also visualizes traffic lights on the map. 
However, these startups do not provide the entire \acp{gui}, including supplemental informational elements, likely due to competitive reasons. 

Both industrial implementations and research on Remote Assistance, especially in \ac{gui} design, remain limited. 
Kettwich et al. \citep{Kettwich2021} designed a remote operator \ac{gui} with six screens and a touchscreen, 
based on a systematic analysis of use cases \citep{Kettwich2020}. 
First, the \ac{gui} was evaluated as a click-dummy by control center professionals, using offline videos. 
Building on this setup, Schrank et al.~\citep{Schrank2024} conducted another study simulating scenarios where an \ac{av} requires Remote Assistance. 
Although the click-dummy achieves good usability and acceptance scores in both studies, it is not designed to determine which informational elements were necessary at which point to achieve these results. 
Similarly, the results from the online user study by Tener and Lanir \citep{Tener2025} allow for general conclusions about their simulation-based click-dummy, revealing good usability scores and relatively positive user acceptance, but lacking information on 
the influence of individual informational elements and the appropriate timing of their display during the teleoperation process.

Consequently, the research questions arise regarding the selection and timing of informational elements within a \ac{gui}
and whether there are distinct differences in the information required for Remote Driving versus Remote Assistance. 
Given the absence of a clearly defined teleoperation process, an initial analysis is necessary to identify the informational elements required at each stage. 
Furthermore, this work presents and evaluates a \ac{gui} prototype that incorporates the informational elements identified as essential. 
This prototype serves as a basis for assessing the effectiveness of the selected elements. 
The detailed methodology is outlined in the following section.

\section{Methodology}
\label{methodology}

The development of the \ac{gui} for remote operation followed the user-centered design process \citep{DIN9241-210}: 1) Understand context of use, 2) Specify user requirements, 3) Design solutions, 4) Evaluate against requirements. 
In the first step, the context of use was defined through interviews, analyzing the teleoperation process for both Remote Driving and Remote Assistance. 
Given the limited user base, primarily within startups developing Remote Driving systems, teleoperation experts were consulted during system development. 
Based on the context of use, the experts assessed the importance of collected informational elements in a second step during the interviews to specify requirements. 
These requirements formed the basis for the \ac{gui} design, which was implemented as a click-dummy in a third step.
In the fourth step, the \ac{gui} was evaluated in an online study. 

The following sections describe the approach used in the expert interview and the online study. 
The interviews and studies were conducted with ethical approval from the ethics committee of the Technical University of Munich (2024-79-NM-BA, 2024-82-NM-BA). 

\subsection{Expert Interview}
\label{expert interview}

To design a display concept for teleoperation, it was first necessary to understand the interaction process between a remote operator and the \ac{av}. 
According to the authors, no publications on the teleoperation process currently exist, so this process was developed through expert interviews. 
Therefore, nine international experts (median age: 33 with $\sigma=8$) were recruited, each with a minimum of one year of teleoperation experience. 

To guide the participants along a generic interaction process, we defined an imaginary situation in which the participant is working as a remote operator responsible for an \ac{av} fleet capable of self-driving but still needs human support in unsolvable situations. 
An \ac{av} requires support by either (A) remotely driving the vehicle or (B) providing Remote Assistance to the vehicle. 
Due to the odd number of participants, five participants were randomly assigned to (A) Remote Driving (P\tsub{RD}) and four participants to (B) Remote Assistance (P\tsub{RA}) without considering the expert's specification. 

\begin{figure}[H]
\centering
\vspace{-0.3cm}
\includegraphics[width = 15.6cm]{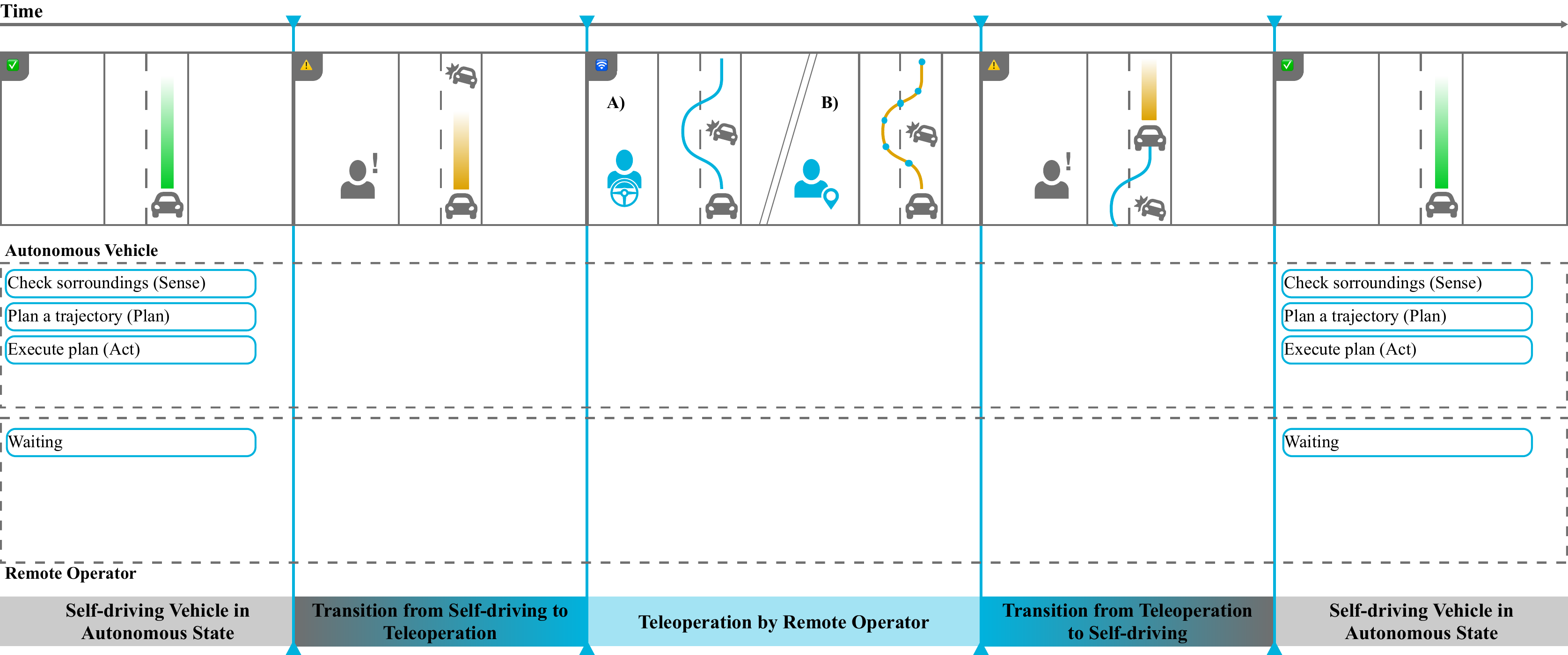}
\caption{Empty Template of the Teleoperation Process with Five Predefined Sections and Predefined Actions in Blue-Bordered Boxes}
\label{fig:template}
\end{figure}

\textbf{Task I} For the first task, the participants were provided with an empty template (Figure~\ref{fig:template}) with five predefined sections from self-driving transitioning to teleoperation and back separated by vertical blue lines. 
The participants' task was to define actions executed by either the \ac{av} or the remote operator during the interaction process. 
Subsequently, these actions should be placed as blue-bordered boxes in one of the dotted boxes related to the \ac{av} and the remote operator. 
The predefined actions in the 1\tsuper{st} and 5\tsuper{th} section could be modified, and additional actions added. 
Multiple actions could be placed in series or in parallel, with different durations represented by the length of the boxes. 
If actions last for multiple predefined sections, they could also be placed across the blue section lines. 
Beyond that, the participants were asked to assess the importance of a standardized teleoperation process for themselves and the industry.


\unskip
\vspace{0.2cm}

In the next step, we wanted to determine which informational element should be shown on a \ac{gui} for the remote operator in which section of the teleoperation process. 
As part of a literature review, more than 60 \acp{gui} in the field of teleoperation and in-vehicle systems were previously analyzed, and their display elements were compiled into a morphological box, resulting in 57 categories of informational components.

\textbf{Task II} For the second task, the experts were asked to evaluate the 57 informational elements. 
The participants were provided with the same template containing the predefined five sections of the teleoperation process (Figure~\ref{fig:template}) and invited to rate the relevance of each informational element within these sections \{0: \textit{irrelevant}, 1: \textit{not necessary, but nice to have}, and 2: \textit{necessary}\}. 
Thereby, the evaluation only assessed the permanent display of information. This task did not cover short-term visuals such as warnings (e.g., for making the remote operator aware of low tire pressure).


\subsection{Online Study}
\label{online study}

\begin{figure}[H]
  \vspace{-0.3cm}
  \centering
  \includegraphics[width = 13.9cm]{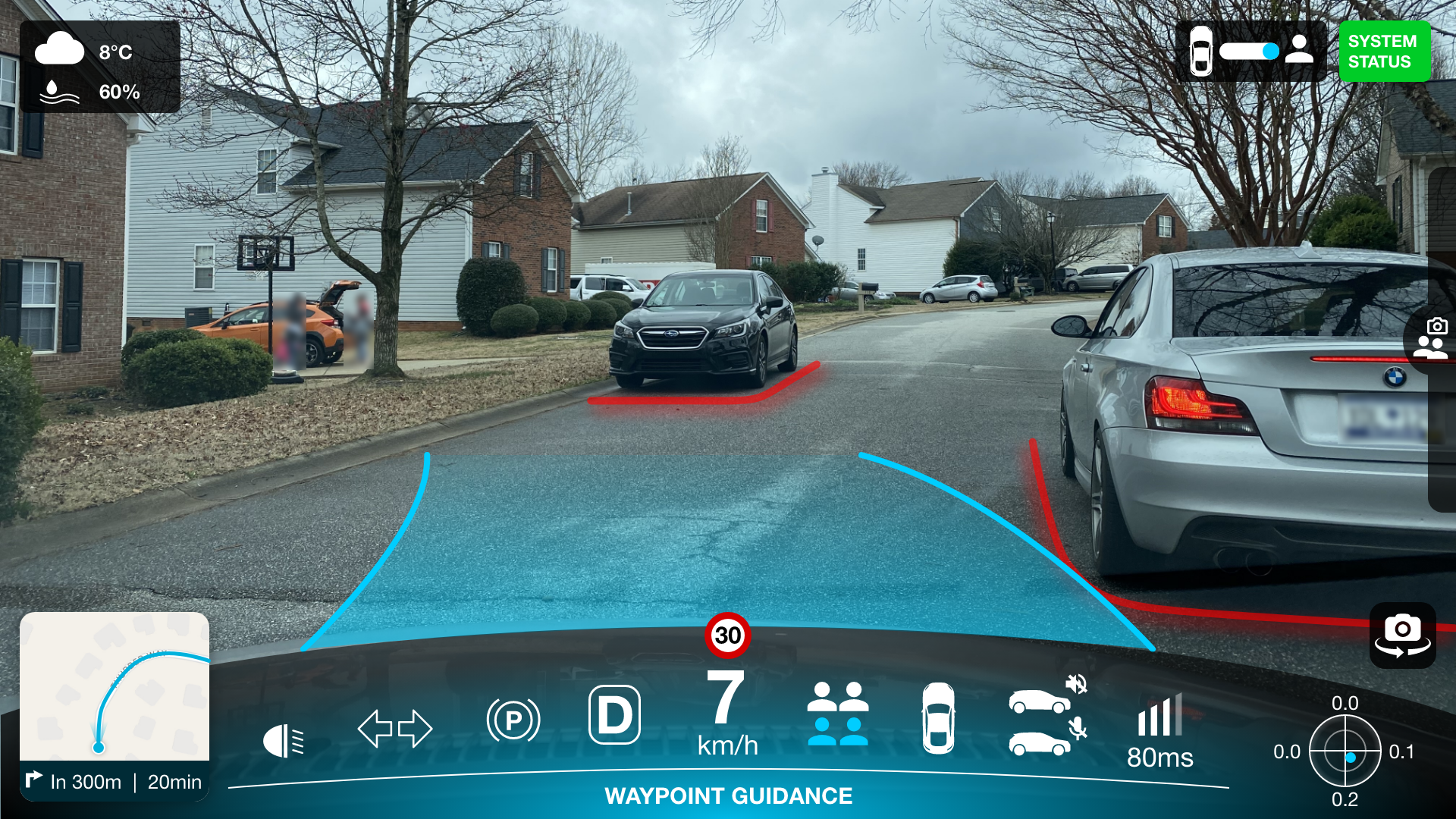}
  \caption{Designed \ac{gui} during Teleoperation Mode (Waypoint Guidance)}
  \label{fig:UIstatic}
\end{figure}

Based on the results of the expert interviews (analyzed in \ref{elements}), we designed a \ac{gui}, shown in Figure~\ref{fig:UIstatic}, containing the informational elements at least classified as 1: \textit{not necessary, but nice to have}. 
The \ac{gui} uses the area of the vehicle's engine hood as a dashboard for displaying vehicle parameters. 
Additional informational elements are overlaid in the top border area of the screen to minimize interference with the video stream. 

\begin{figure}[H]
  \vspace{-0.1cm}
  \centering
  \includegraphics[width=13.9cm]{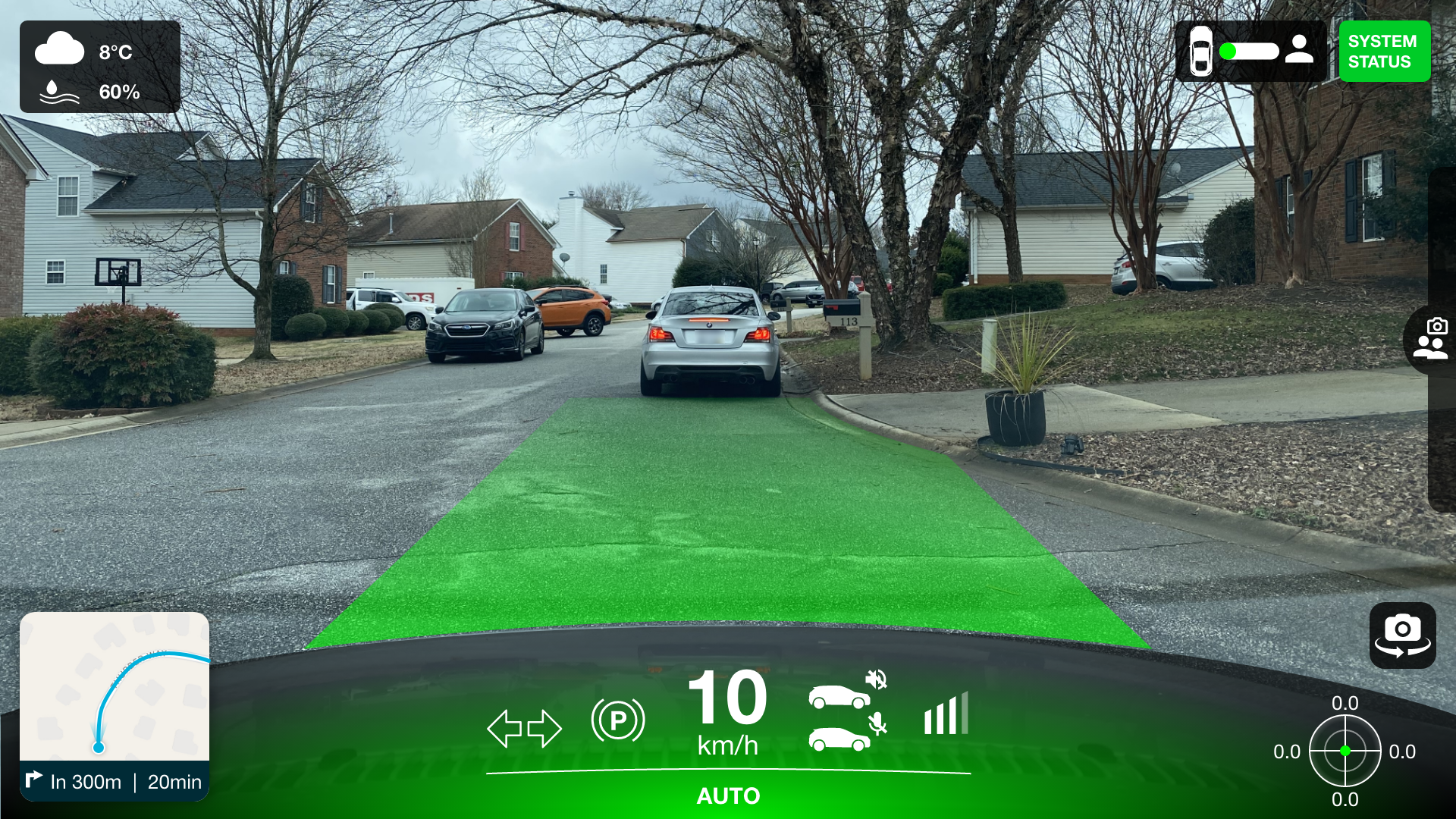}
  \caption{Designed \ac{gui} Variant with Reduced Set of Display Elements during Autonomous Self-Driving for Dynamic Adaptation across the Teleoperation Process} 
  \label{fig:UIdynamic}
\end{figure}

For the remote monitoring during the self-driving in the 
transition sections, two display alternatives with varying levels of information density were developed. 
The first display variant corresponds to the \ac{gui} in teleoperation mode (Figure~\ref{fig:UIstatic}), ensuring a static \ac{gui} throughout the teleoperation process. 
The second display variant (Figure~\ref{fig:UIdynamic}) features a reduced number of elements tailored to the teleoperation mode, allowing the \ac{gui} to dynamically adapt across the stages of the teleoperation process, incorporating the results of the expert interview. 
When switching to teleoperation, both \ac{gui} variants change their appearance from a green to a blue tint, indicating that human input is required.

The online study aimed to evaluate and compare the static and dynamic \ac{gui} variants while validating the experts' assessment of the necessity of displaying certain informational elements during teleoperation. 
To simulate real-world teleoperation, we developed an interactive online prototype that participants could click through, enabling a remote study without requiring peripherals like pedals or steering wheels by focusing solely on Remote Assistance. 
Therefore, we took a series of photos of a scenario where the ego vehicle was driving about 50\,m through an American neighborhood. 
To proceed to the next interactive screen and move the vehicle forward, the participants had to click on the road in the picture. 
Otherwise, the \ac{gui} showed an error message. 

\textbf{Task I} After an introduction to teleoperation, the participants were required to interact with the static \ac{gui} while paying attention to the display elements. 
After starting the first interactive screen by clicking a button, the participant monitored the \ac{av} in autonomous mode until the \ac{av} requested the participant to support it. 
The participant must guide the \ac{av} through the scenario by clicking a waypoint in the desired direction on the street surface. 
After solving the scenario, the participant was informed that the \ac{av} can continue driving autonomously. 
Hence, the control was handed over to the \ac{av}. 
After the interaction with the prototype, the participants evaluated the static \ac{gui} using the \ac{sus} \citep{Bangor2009} and the short \ac{ueq} \citep{UEQ}.

\textbf{Task II} In the second task, the participants were introduced to the dynamic \ac{gui} and had to solve the same scenario with this display variant again. We did not make any additional changes except for the reduced set of display elements during autonomous driving mode. 
After completing the \ac{sus} \citep{Bangor2009} and \ac{ueq} \citep{UEQ} again, the participants were asked to identify their preferred \ac{gui} option and offer suggestions for potential improvements.

\textbf{Task III} Finally, the participants should do a card sorting by allocating every informational element of the \ac{gui} to one of the following categories \{0: \textit{irrelevant}, 1: \textit{not necessary, but nice to have}, and 2: \textit{necessary}\}. 
Thereby, every informational element was represented by a card containing the icon and a textual designation. 

\textbf{Participants} For the online study, a total of $N=36$ participants (4 female, 32 male) with an average age of 26.3 years ($SD=6.0$) were recruited. 
Every participant except for one had a valid driver's license. 
Of these, 28 participants stated gathering driving experience for 10 or fewer years, seven participants for 11 to 20 years, and one for more than 30 years. 
14 participants drove almost every day, five a few days a week, 12 a few days a month, and four a few times a year. 
31 participants stated doing only short-distance travel ($<$\,200\,km per round trip), three participants do middle-distance travel (201\,km to 500\,km per round trip) and one participant does long-term travel ($>$\,500\,km per round trip).
The participants' average \ac{ATI} (short version) score is $M=4.42$ with $SD=1.08$, and $\alpha=0.89$.

\section{Results}
\label{results}

The following section summarizes the results from the expert interviews on the teleoperation process and the evaluation of informational elements, as well as the results from the online study assessing the designed \ac{gui} as a click-dummy. 

\subsection{Teleoperation Process}
\label{process}

Before designing the \ac{gui}, the first step was to conduct an expert interview to define the context of use. 
Each expert, assisted by the interviewer, specified a teleoperation process in the given template (Figure~\ref{fig:template}) by defining different actions for the \ac{av} or remote operator. 
Since the experts did not modify the self-driving sections s1 and s5, we only focus on the Transition from Self-driving to Teleoperation (s2), Teleoperation by Remote Operator (s3), and the Transition from Teleoperation to Self-driving (s4). 
Similar actions were grouped wherever possible to compare the resulting process steps. 
Tasks executed by the \ac{av} are plotted in orange, and those executed by the remote operator in purple. 
The more often the experts mentioned a certain action, the higher the number and the more intense the color.

For Remote Driving (Table~\ref{table:expert_task_1_RD}), the main process starts with the \ac{av} driving autonomously and continuously sending situation data while the remote operator is waiting. 
However, an expert emphasized that teleoperation is only profitable if waiting time is utilized for other tasks like fleet monitoring. 
Once the \ac{av} recognizes a problem, it sends a notification to the remote operator while executing a minimal risk maneuver. 
P2\tsub{RD} suggested that the optimal take-over request design enables the vehicle to ask for support.
After receiving the notification, the remote operator assesses the situation. 
If the \ac{av} reached a safe state, the remote operator takes over for teleoperation and chooses one of the interaction concepts offered by the \ac{av} that the \ac{av} has to approve. 
Thus, the remote operator sends driving commands, which the \ac{av} monitors and executes consecutively, whereby the \ac{av} can offer the remote operator assistance or automatically intervene if necessary. 
In turn, the \ac{av}'s execution is monitored by the remote operator. 
Once the difficult situation is bypassed, the remote operator guides the \ac{av} back into a safe state and requests autonomous driving. P3\tsub{RD} emphasized that the vehicle must inform the remote operator that self-driving is available again. Therefore, the \ac{av} checks the confidence for self-driving, approves the request, and transitions back to autonomous driving. Afterward, the remote operator monitors the \ac{av} until both disconnect.

\begin{table*}[!htbp]
    \caption{Specified Teleoperation Process Steps for Remote Driving, Including Number of Mentions
    \label{table:expert_task_1_RD}
    }
      \centering
     \includegraphics[trim=0cm 4.3cm 0cm 0cm, clip, width = 15.6cm]{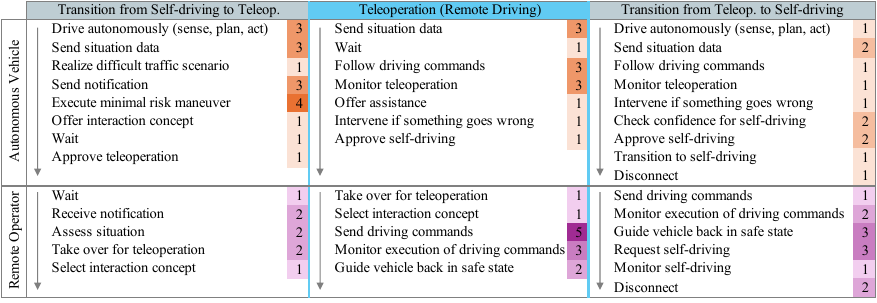}
    \vspace{-0.2cm}
\end{table*}

For Remote Assistance (Table~\ref{table:expert_task_1_RA}), the experts described the process during the Transition from Self-Driving to Teleoperation (s2) similar to the Remote Driving procedure. 
However, the interaction is continuously supported by ADAS systems while the system collects information for learning.
Furthermore, the experts emphasize that the remote operator first connects to the \ac{av} to assess the situation when a notification from the \ac{av} is received, checked, and confirmed. 
If the \ac{av} reaches a safe state, the remote operator can request or accept teleoperation and takes over. 
In the specific scenario of Waypoint Guidance \citep{Maj2022}, the remote operator plans a trajectory by setting a series of waypoints. 
P4\tsub{RA} proposes that this trajectory can also include a point where the transition to autonomous driving is performed automatically. 
Once the remote operator confirms the final trajectory, they monitor the \ac{av}'s execution. 
Thereby, the \ac{av} checks if it is back on the originally planned trajectory and, if so, transitions back to self-driving. The remote operator can intervene in this transition if something goes wrong. 
If the transition to autonomous driving is not done automatically, the remote operator tries to stop the vehicle and manually hands over to the \ac{av}. 
As a final step, the \ac{av} disconnects from the remote operator. 

\begin{table*}[!htbp]
      \centering
      \caption{Specified Teleoperation Process Steps for Remote Assistance, Including Number of Mentions
    \label{table:expert_task_1_RA}
    }
    \includegraphics[trim=0cm 0cm 0cm 5.2cm, clip, width = 15.6cm]{figures/expert_task_1_teleoperation_process_thick_lines.pdf}
    \raggedright
    \noindent{\footnotesize{* The steps highlighted in gray are optional and only required when automated handover is not possible, necessitating a manual transition from a stationary position.}}
    \vspace{-0.2cm}
\end{table*}

In the post-task survey, the question "How important is a standardized teleoperation process for you?" was answered with $4.6$ on average on a 5-point Likert scale (1: \textit{very unimportant}, 5: \textit{very important}). 
The importance of a standardized teleoperation process for the industry was assessed at $4.7$. 
However, P1\tsub{RA} warned that overly detailed governmental standardization may hinder development, while P3\tsub{RA} noted that higher standardization improves efficiency but is challenging due to varying conditions. 
P2\tsub{RD} raised key research questions, including the optimal operator-to-vehicle ratio, minimum latency, expected market penetration of Level 5 autonomy, and the benefits of Remote Driving over Assistance.

\subsection{Informational Elements}
\label{elements}

The experts rated each collected informational element on a scale of $\{0, 1, 2\}$, regarding their assigned teleoperation concept (Remote Driving or Remote Assistance) and the respective section (s1-s5) in the teleoperation process. The average results are visualized in Table~\ref{table:expert_task_2} with darker shades and higher values indicating greater relevance of an element.

\begin{table*}[htbp]
    \centering
    \caption{Heat Map for Remote Assistance, Remote Driving and their Comparison based on the Experts’ Evaluation of 57 Informational Elements Identified through Literature Review \newline \{0: \textit{irrelevant}, 1: \textit{not necessary, but nice to have}, and 2: \textit{necessary}\} \newline (x) marks elements intended for the \ac{gui} but omitted due to scenario and click-dummy limitations}
    \label{table:expert_task_2}
    \vspace{-0.1cm}
    \includegraphics[trim=1.8cm 6.3cm 2cm 2cm, clip, width = 17.0cm]{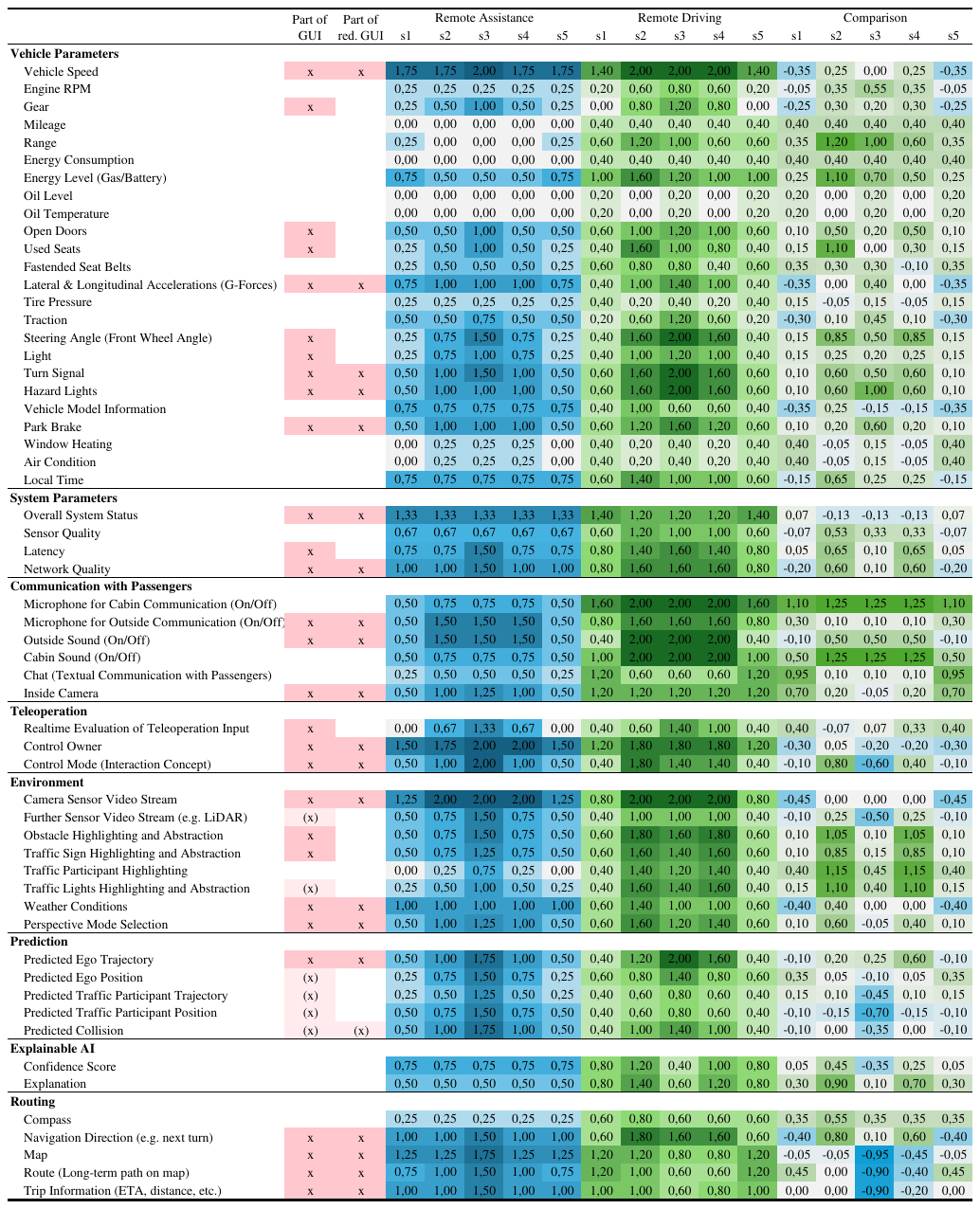}
    \raggedright
    \noindent{\footnotesize{* Comparison: positive numbers indicate 
    more importance for Remote Driving, negative numbers for Remote Assistance\\
    ** s1: Self-driving Vehicle in Autonomous State, s2: Transition from Self-driving to Teleoperation, s3: Teleoperation by Remote Operator, s4: Transition from Teleoperation to Self-Driving, s5: Self-driving Vehicle in Autonomous State}}
\end{table*}

For Remote Assistance the experts consider informational elements like Engine RPM, Mileage, Range, Energy Consumption and Level, Oil Level and Temperature, as well as a Compass as irrelevant, regardless of the teleoperation process section (s1-s5). In contrast, elements like Vehicle Speed, Control Owner and Mode, as well as a Camera Sensor Video Stream, appear to be mostly necessary at all times. 

Although the Remote Driving heat map is comparable, it shows more high-necessity ratings (value $2.00$), emphasizing the need for additional informational elements like Steering Angle, Turn Signal that go hand in hand with Hazard Lights as well as a Microphone for Cabin Communication, Outside Sound, Cabin Sound, and the Predicted Ego Trajectory. 
However, P1\tsub{RA}, P2\tsub{RD}, and P3\tsub{RD} noted that the need to toggle cabin or outside sound can be eliminated by setting up a system that only forwards relevant conversations to the remote operator. 
Furthermore, a few ratings, like Energy Level, Range, Used Seats, Vehicle Model Information, and Local Time, break the symmetry and are most important during the transition to Remote Driving (s2), as the remote operator may need this information to gain sufficient situational awareness.

The \textit{Comparison} heat map compares both evaluations for Remote Assistance and Remote Driving. 
The cells' values within $[-2.0, 2.0]$ were calculated by subtracting the Remote Assistance from the Remote Driving values. 
Hence, positive numbers (marked in green) indicate informational elements that were more important for Remote Driving. Vise versa, negative numbers (marked in blue) indicate elements considered more important for Remote Assistance. 
Notably, Map, Route, and Trip Information were rated as more critical for Remote Assistance, whereas Traffic Signs, Lights, Participant, and Object Highlighting and Abstraction were considered more relevant for Remote Driving.

Overall, expert assessments of Remote Assistance and Remote Driving concepts do not differ significantly. However, the predominantly green shading in the \textit{Comparison} column indicates that many informational elements were considered slightly more important for Remote Driving ($0.22$ higher on average compared to Remote Assistance, Table~\ref{table:expert_task_2}). 

\subsection{Online Study}
\label{results online study}

\textbf{Design Derivation} 
Since the online study focuses on evaluating a novel \ac{gui} for Remote Assistance, the \ac{gui} concept was derived based on the results of the expert interviews for Remote Assistance, particularly considering the highest ratings of the elements within the teleoperation section s3 for the static \ac{gui} variant and the importance towards all sections for the reduced \ac{gui} (Table~\ref{table:expert_task_2}). 
The derived design for the click-dummy (Figure~\ref{fig:UIstatic}) only contains the informational elements that were at least classified as \textit{nice to have, but not necessary} by every expert, i.e. an average value of at least $1.00$. 

However, Traffic Lights Highlighting and Abstraction were excluded because the scenario in the online study does not include any traffic lights but would be applied analogously to the traffic signs in the \ac{gui}. 
As the scenario also does not include moving participants, no Predicted Traffic Participant Trajectory or Position was highlighted. 
The explicit display of the Further Sensor Video Stream (e.g., Lidar) and Predicted Collisions was omitted, as detected objects on the road are already highlighted with red line markings, and no collisions were possible due to the fixed scenario screens to click through. 
Similarly, the Predicted Ego Position can be inferred from the Predicted Ego Trajectory, visualized as a driving corridor. 
The resulting set of informational elements is indicated with an \textit{x} and marked red in column \textit{Part of \ac{gui}} in Table~\ref{table:expert_task_2}. All elements omitted for the reasons mentioned above are marked with \textit{(x)}. 
The reduced \ac{gui} (Figure~\ref{fig:UIdynamic}) contains fewer informational elements during monitoring autonomous self-driving that were, on average, considered as 1: \textit{nice to have, but not necessary} in other sections than teleoperation (s1, s2, s4, s5). 

\textbf{Evaluation} The online study aimed to evaluate and compare the developed static and dynamic \ac{gui}. 
Since waypoints could only be placed on the road surface, it was possible to track misclicks. 
Misclicks occurred 10 times while using the \guis and 14 times with \guid (Table~\ref{table:online_study_results}). 


\begin{table}[!htbp]
\centering
\caption{Comparison Results of the Static and Dynamic \ac{gui}}
\isPreprints{\centering}{} 
\centering
\begin{tabular}[t]{lrrrrrr}
\toprule
& \multicolumn{3}{c}{\guis} & \multicolumn{3}{c}{\guid} \\
& $Mdn$ & $M$ & $SD$ & $Mdn$ & $M$ & $SD$ \\
\midrule
Misclicks & 10 & -- & -- & 14 & -- & -- \\
Task Completion Time (s) & 61.73 & 76.7 & 37.7 & 41.25 & 64.8 & 69.6 \\
\ac{sus} Score & 70.00 & 68.5 & 12.8 & 78.75 & 76.6 & 12.5 \\
\ac{ueq} Score & 0.56 & 0.41 & 0.90 & 0.63 & 0.67 & 0.92 \\
\hspace{5mm}Pragmatic Quality & 0.88 & 0.65 & 1.08 & 1.13 & 0.94 & 1.11 \\
\hspace{5mm}Hedonic Quality & 0.38 & 0.16 & 1.10 & 0.50 & 0.40 & 1.08 \\
\bottomrule
\end{tabular}
\label{table:online_study_results}
\end{table}

Guiding the vehicle via clicks through the situation using \guis took an average of $76.7\,s$ 
whereas \guid reduced this time by $16\,\%$. 
As the Saphiro-Wilk test shows no normal distribution for the Task Completion Time, the Wilcoxon test was applied to assess significant differences.
The results reveal that using the \guid was significantly faster 
than using the \guis 
($z = -3.97$, $p = .00007$) with an effect size of $r = .66$, indicating a strong effect according to Cohen \citep{Cohen1992}.

The participants rated the usability of \guis on average with a \ac{sus} score of $68.5$, 
putting it within the higher “marginal” acceptability range and labeling it with the adjective "ok" \citep{Bangor2009}.
In comparison, the \guid received a \ac{sus} score of $76.6$, 
ranking as "acceptable" while shifting the adjective to "good". 
Due to the lack of normal distribution, a Wilcoxon test was conducted for the \ac{sus} scores, revealing that the \ac{sus} score for the \guid was significantly higher 
compared to the \guis 
($z = -4.11$, 
$p = .00004$) with an effect size of $r = .69$, indicating a strong effect \citep{Cohen1992}. 

The user experience of the \guis was rated lower compared to the \guid in pragmatic, hedonic, and overall quality within the neutral evaluation range (Table~\ref{table:online_study_results}).
The \guid scored the highest rated value 
in pragmatic quality, reaching a positive evaluation. 
Although the \ac{ueq} data is normally distributed, the t-test showed no significant difference in the overall \ac{ueq} score ($t = -2.007$, $p = .0525$), 
but there is a significant difference ($t = -2.160$, $p = .0377$) for the Pragmatic Quality between \guid 
and \guis 
with a small practical effect ($r = .36$) \citep{Cohen1992}. 
For the Hedonic Quality, the t-test shows no significant difference ($t = -1.355$, $p = .1842$). 

In the post-task survey, users rated their preference between \guis and \guid: 4 favored \guis, 11 preferred \guid, 9 liked both, and 12 saw no difference.

\textbf{Informational Elements} 
The result of the card sorting is visualized in Table~\ref{table:card_sorting}. 

\begin{table}[htbp]
      \centering
      \vspace{-0.2cm}
      \caption{Comparison of the Experts’and Participants’ Rating of Informational Elements during Remote Assistance and their Difference: Positive Values Indicate Higher Participant Ratings (Pink), Negative Values Higher Expert Ratings (Turquoise) \newline \{0: \textit{irrelevant}, 1: \textit{not necessary, but nice to have}, and 2: \textit{necessary}\}}
    \label{table:card_sorting}
    \isPreprints{\centering}{} 
    \includegraphics[trim=1.8cm 6.2cm 3.15cm 2cm, clip, width = 11.0cm]{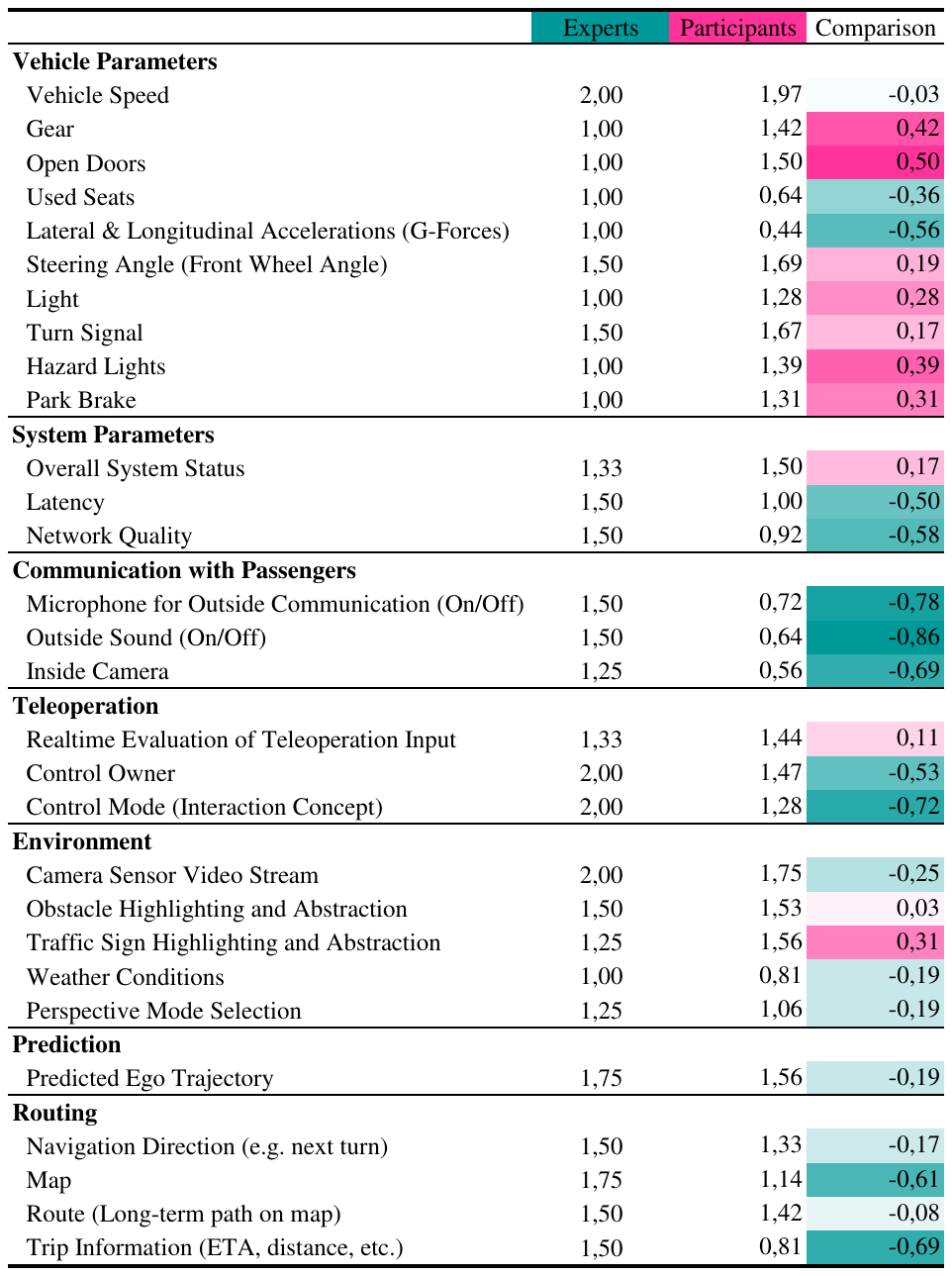}
\end{table}

The \textit{Experts} and \textit{Participants} columns contain the respective averaged ratings of each informational element during the teleoperation section (s3) for Remote Assistance, while the \textit{Comparison} column highlights their differences. 
Positive values (marked in pink) indicate elements rated higher by participants, whereas negative values (marked in turquoise) signify elements rated higher by experts. 
The majority of vehicle parameters were rated significantly higher by the participants. 
On the other side, informational elements of system parameters and communication with passengers, teleoperation, and routing were rated notably higher by the experts. 
Overall, four participants claimed in the suggestions for improvement that the \acp{gui} in general contain too much information for the task. 
Three participants criticized the prototype's limited interaction and the resulting challenges. 
Furthermore, three participants expressed a preference for the ability to customize the \ac{gui} to better suit their individual needs rather than relying on a default configuration. 

\section{Discussion and Limitations}
\label{discussion}

Following the results, this section discusses their outcomes and limitations to lay the foundation for future research and ensure the proper interpretation of the findings.

\subsection{Teleoperation Process}
\label{dis:teleop_process}

The following general process steps are derived from the Remote Driving and Remote Assistance process defined during the expert interviews (Table~\ref{table:expert_task_1_RD} and~\ref{table:expert_task_1_RA}), assuming the remote operator is not yet in the loop. It should be noted that 
the experts may not have explicitly stated certain aspects, as they were presumed to be self-evident, such as the \ac{av} also transmitting situation data to the remote operator in Remote Assistance.
\begin{enumerate}
    \item \ac{av} recognizes complex scenario or problem
    \item \ac{av} notifies the remote operator while executing a minimal risk maneuver 
    \item Remote operator receives notification, connects to \ac{av} and assesses situation
    \item If the \ac{av} reached a safe state, the remote operator takes over for teleoperation
    \item Thereby, the \ac{av} can offer different teleoperation concepts
    \item Remote operator chooses a teleoperation concept (Remote Driving or Remote Assistance)
    \item \ac{av} has to approve the teleoperation with the chosen interaction concept
    \item \textit{Execution of the dynamic driving task through the selected teleoperation concept (specific individual steps)}
    \item Thereby, the \ac{av} checks and approves confidence for self-driving
    \item Handing over to self-driving upon request of the remote operator or offer from the \ac{av}, or through automated transfer (especially in Remote Assistance)
    \item Remote operator monitors self-driving and can intervene if something goes wrong
    \item Remote operator and \ac{av} disconnect
\end{enumerate}
Step 8) is further divided into substeps, with Remote Driving involving more low-level commands, whereas Remote Assistance remains high-level due to the continued presence of autonomous features. 
This would allow an automatic transition back to self-driving for Remote Assistance. 
However, the handshakes in step 4 and 10 remain unclear. 

If the remote operator is already connected to the \ac{av} and assessing the situation, they can request teleoperation and take over according to step 4). 
The question remains: Does the remote operator request teleoperation intervention manually, or does the \ac{av} enter teleoperation mode autonomously and await commands? 
Similarly, step 10) involves the second handover transitioning back to self-driving. 
Further research should be conducted to determine which handover option is promising regarding transition time and reliability, as well as if and how an AV could automatically switch back to autonomous driving. 

A further field of research is the right timing of a handover. 
Typically, the vehicle will be handed over to the remote operator from a minimal risk state, which is likely to be a standstill. 
However, a transition process that enables seamless switching between autonomous driving and teleoperation without a minimal risk maneuver is still unexplored. 
Likewise, the transition to self-driving can occur either from a standstill or while 
in motion.

Overall, experts generally consider standardizing the teleoperation process to be important to very important. 
Nevertheless, standardization should not hinder development but allow adaptability across different applications and conditions. 
Variations in the interaction, such as the handover to self-driving (step 10), present significant challenges for establishing a standardized teleoperation process.

\subsection{Informational Elements}
\label{dis:display_elements}

The evaluation of informational elements reduced the number of collected potential display components by more than half, minimizing the \ac{gui} and ensuring the remote operator's attention remains focused on essential information.

The initial experts' assessment of informational elements indicates that features such as Maps, Route, and Trip Information are considered more critical for Remote Assistance, aligning with existing applications in Section~\ref{introduction}. 
In contrast, Traffic Signs, Traffic Lights, Participant and Object Highlighting and Abstraction were rated as more important for Remote Driving, likely because the remote operator is directly responsible for executing the driving task and maintaining vehicle control. 
Therefore, the remote operator may need to gain a more detailed situational awareness than in Remote Assistance. 
The overall higher rated relevance of informational elements in Remote Driving also suggests an increased need for situational awareness in this teleoperation mode. 


However, it is essential to note that the experts came from various fields of teleoperation applications. 
Since they were randomly assigned to the Remote Driving and Remote Assistance groups, it's possible that one sector was overrepresented in a group. 
This may explain why in Remote Driving elements for communication with passengers were considered more important, likely due to the higher proportion of experts from the passenger transport sector compared to the Remote Assistance group, which may have had more experts from logistics applications. 
This effect can be mitigated by increasing the participant pool or evenly distributing the experts based on their prior experience. 

Despite the small number of experts ($N=9$) and random distribution, the relatively symmetrical color distribution in the heat map in Table~\ref{table:expert_task_2} reflects that the elements are perceived as equally necessary or unnecessary, whether for Remote Driving or Assistance. 

Comparing the expert's rating with the results from the participants of the online study (Table~\ref{table:card_sorting}), the assessments for Remote Assistance can largely be confirmed by the larger participant sample. 
However, participants rated vehicle parameters as more important than experts, whereas experts placed greater emphasis on elements related to Communication with Passengers, Map, Trip Information, Latency and Network Quality indicators, as well as Control Owner and Mode display. 
This may be because participants in the online study did not have to interact with these elements, making the information less relevant in the click-dummy scenario without an actual driving task. 
It also shows that the experts take a different perspective compared to the laypersons due to their prior experience. 

Overall, it was noted in the online study that the \ac{gui} was overloaded with information and should be further reduced. 
However, it is again crucial to consider that many elements were not needed in the click task without a vehicle connection. 
In the next step, a user study involving the execution of teleoperation using a real-world vehicle should be conducted. 

In the future, it is conceivable that the informational elements could be individually arranged and customized by the remote operator, as suggested by three participants.

\subsection{Online Study}
\label{dis:online_study}

The expert interview findings were translated into a \ac{gui} prototype, implemented in a static and dynamic variant using a click-dummy. 
However, one-third of participants reported not noticing any difference between the two versions. 
Still, the dynamic \ac{gui} received a significantly higher rating in terms of usability. 
This result may be influenced by the presentation order, as participants were always shown the static \ac{gui} first, followed by the dynamic version. 
Consequently, increased familiarity with the system in the second trial could have led to better evaluations.

Similarly, the faster task completion times observed with the dynamic \ac{gui} may be attributed to the fixed order of presentation and potential learning effects. 
Additionally, both \ac{gui} variants were tested using the same scenario, which may have amplified the impact but led to better comparability.

The number of misclicks can likely be attributed to participants' curiosity, as they attempted to challenge the system and explore its limitations. 
Nevertheless, the \ac{gui} variants achieved a higher marginal acceptable to good \ac{sus} score, with pragmatic quality in the user experience falling within the (almost) green range. 
In contrast, the hedonic quality received lower ratings, remaining within a neutral range, possibly due to the \ac{gui} being evaluated only in a prototypical click-dummy without real vehicle integration. 
Three participants even criticized the lack of proper interaction. 
Therefore, the results of usability and user experience cannot be solely assigned to the \ac{gui} but are also influenced by the interaction with the prototype. 
Future studies should explore the combined impact of interaction and display concepts on usability and system performance, considering various interaction concepts and real-world applications.  
Thereby, studies should be conducted under real-world conditions, such as video streaming with associated transmission latencies and interaction with a vehicle.

In summary, 30\,\% of the participants preferred the dynamic \ac{gui} (33,3\,\% noticed no difference, 25\,\% liked both), which was generally rated higher regarding usability and user experience. 
Therefore, the display should adapt throughout the teleoperation process, providing only the necessary elements at each stage. 
However, the dynamic \ac{gui} and its reduced display during the automated driving monitoring phase still need to be further investigated with a real vehicle connection. 
In particular, the remote operator notification, including the switch to the scenario, and the subsequent handover to teleoperation require further evaluation.

\section{Conclusions}

As the core of this work, a Graphical User Interface (\ac{gui}) for the teleoperation of \acp{av} was developed following the user-centered design process. 
First, process steps were defined by experts, outlining tasks to be performed by the \ac{av} or remote operator. 
Thereby, a major unresolved issue was the handshake to the remote operator and back to self-driving. 
Based on these steps, the experts evaluated the necessity of informational elements gathered from the state of the art. 
Building on the results, a click-dummy was created, and the necessity of the informational elements was reassessed in an online study with a larger sample. 
Additionally, a comparison was made between a static and a dynamic \ac{gui} showing reduced display elements during the automation monitoring phase. 
Although one-third of the participants reported no noticeable difference between the \acp{gui}, the usability of the dynamic \ac{gui} was rated significantly higher. Overall, both \acp{gui} received good System Usability Scale (\ac{sus}) ratings. 
However, the User Experience Questionnaire (\ac{ueq}) score suggests that there is still potential for improvement in terms of user experience. 
Task Completion Time was also significantly faster with the dynamic \ac{gui}. 
Based on the expanded insights, 
the \ac{gui} design should be further refined and integrated into existing interaction concepts. 
This enables iterative development, including future studies involving interaction with a vehicle.
Overall, the developed display concept could serve as a standard interface for presenting key information in teleoperation research.




\vspace{6pt} 





\authorcontributions{
As the first author, Maria-Magdalena Wolf initiated the idea of this paper and contributed essentially to its conception and content. 
Henrik Schmidt created the click-dummy, executed and analyzed the expert interviews and online study, and contributed to the writing process. 
Michael Christl supported the statistical analysis of the online study. 
Jana Fank critically revised the paper and contributed through insightful discussions. 
Frank Diermeyer made essential contributions to the conception of the research project and revised the paper critically. 


Conceptualization, M.-M.W.; methodology, M.-M.W.; software, H.S.; validation, M.-M.W.; formal analysis, H.S and M.C.; investigation, H.S.; resources, M.-M.W. and H.S.; data curation, H.S. and M.-M.W.; writing---original draft preparation, M.-M.W. and H.S.; writing---review and editing, J.F. and F.D.; visualization, H.S. and M.-M.W.; supervision, M.-M.W.; project administration, M.-M.W. and F.D.; funding acquisition, F.D. All authors have read and agreed to the published version of the manuscript.
} 


\funding{This research was funded by the Federal Ministry of Economic Affairs and Climate Action of Germany~(BMWK) within the project Safestream~(FKZ~01ME21007B).}

\institutionalreview{The studies were conducted in accordance with the Declaration of Helsinki, and approved by the Ethics Committee of the Technical University of Munich (2024-79-NM-BA 02.08.2024; 2024-82-NM-BA 04.09.2024).}


\informedconsent{Informed consent was obtained from all subjects involved in the study.}


\dataavailability{The original contributions presented in these studies are included in the article. Further inquiries can be directed to the corresponding author.}

\conflictsofinterest{The authors declare no conflicts of interest. The funders had no role in the design of the study; in the collection, analyses, or interpretation of data; in the writing of the manuscript; or in the decision to publish the results.}




\abbreviations{Abbreviations}{
The following abbreviations are used in this manuscript:
\\

\noindent 
\begin{tabular}{@{}ll}
ATI & Affinity for Technology Interaction\\
AV & Automated Vehicle\\
GUI & Graphical User Interface\\
SUS & System Usability Scale\\
UEQ & User Experience Questionnaire
\end{tabular}
}





\reftitle{References}


 \bibliography{bibliography}

\begin{thebibliography}{999}

\bibitem[{Council of European Union}(2022)]{eu-1426-2022}
{Council of European Union}.
\newblock Council Regulation (EU) No 1426/2022,  2022.

\bibitem[LLC(2025)]{Waymo2025}
LLC, W.
\newblock Waymo - Self-Driving Cars - Autonomous Vehicles - Ride-Hail.
\newblock Available online: \url{https://waymo.com/} (accessed on 30.01.2025).

\bibitem[Cru(2025)]{Cruise2025}
Cruise Driverless Rides | Autonomous Vehicles | Self-Driving.
\newblock Available online: \url{https://www.getcruise.com/} (accessed on
  30.01.2025).

\bibitem[Jin(2021)]{Jin2021}
Jin, H.
\newblock Insight: A secret weapon for self-driving car startups: Humans.
\newblock Available online:
  \url{https://www.reuters.com/business/autos-transportation/secret-weapon-self-driving-car-startups-humans-2021-08-23/}
  (accessed on 2025-01-29).

\bibitem[Kelkar et~al.(2025)Kelkar, Heineke, and Kellner]{Kelkar2025}
Kelkar, A.K.; Heineke, K.; Kellner, Martin with~Smith, A.S.
\newblock Remote-driving services: The next disruption in mobility innovation?
  {\bf 2025}.

\bibitem[Valeo(2023)]{Valeo2023}
Valeo.
\newblock Valeo to showcase major innovations at iaa mobility 2023.
\newblock Available online:
  \url{https://www.valeo.com/en/valeo-to-showcase-major-innovations-at-iaa-mobility-2023/}
  (accessed on 26.1.2025).

\bibitem[Dri(2025)]{DriveU2025}
Teleoperation in all use cases and all levels of autonomy.
\newblock Available online: \url{https://driveu.auto/} (accessed on
  26.01.2025).

\bibitem[Einride(2024)]{Einride2024}
Einride.
\newblock Balancing humanity and autonomy - The Einride Remote Interface allows
  operators to monitor a fleet of vehicles and keep an eye on their progress.
\newblock Available online:
  \url{https://www.einride.tech/what-we-do/autonomy{\#}automate} (accessed on
  26.01.2025).

\bibitem[Fernride(2024)]{Fernride2024}
Fernride.
\newblock Human Assisted Autonomy.
\newblock Available online: \url{https://www.fernride.com/system} (accessed on
  12.04.2025).

\bibitem[Mutzenich et~al.(2021)Mutzenich, Durant, Helman, and
  Dalton]{Mutzenich2021}
Mutzenich, C.; Durant, S.; Helman, S.; Dalton, P.
\newblock Updating our understanding of situation awareness in relation to
  remote operators of autonomous vehicles.
\newblock {\em Cognitive Research: Principles and Implications} {\bf 2021},
  {\em 6},~9.
\newblock {\url{https://doi.org/10.1186/s41235-021-00271-8}}.

\bibitem[Carsten and HF-IRADS(2020)]{Carsten2020}
Carsten, O.M.J.; HF-IRADS.
\newblock Human Factors Challenges of Remote Support and Control A Position
  Paper from HF-IRADS1 1. Introduction.
\newblock  2020.

\bibitem[Colavita(1974)]{Colavita1974}
Colavita, F.B.
\newblock Human sensory dominance.
\newblock {\em Perception and Psychophysics} {\bf 1974}, {\em 16},~409–412.
\newblock {\url{https://doi.org/10.3758/bf03203962}}.

\bibitem[Georg et~al.(2020)Georg, Putz, and Diermeyer]{Geo2020b}
Georg, J.M.; Putz, E.; Diermeyer, F.
\newblock Longtime Effects of Videoquality, Videocanvases and Displays on
  Situation Awareness during Teleoperation of Automated Vehicles.
\newblock In Proceedings of the 2020 IEEE International Conference on Systems,
  Man, and Cybernetics (SMC),  2020, pp. 248--255.
\newblock {\url{https://doi.org/10.1109/SMC42975.2020.9283364}}.

\bibitem[Boker and Lanir(2023)]{Boker2023}
Boker, A.; Lanir, J.
\newblock Bird’s Eye View Effect on Situational Awareness in Remote Driving.
\newblock In Proceedings of the Adjunct Proceedings of the 15th International
  Conference on Automotive User Interfaces and Interactive Vehicular
  Applications. ACM,  September 2023, AutomotiveUI ’23.
\newblock {\url{https://doi.org/10.1145/3581961.3609878}}.

\bibitem[Voysys(2020)]{Voysys2020}
Voysys.
\newblock Field of view affects sense of speed.
\newblock Available online:
  \url{https://www.youtube.com/watch?v=1D3j352{\_}jsM} (accessed on
  2025-04-10).

\bibitem[Cabrall et~al.(2019)Cabrall, Stapel, Besemer, Jongbloed, Knipscheer,
  Lottman, Oomkens, and Rutten]{Cab2019}
Cabrall, C.; Stapel, J.; Besemer, P.; Jongbloed, K.; Knipscheer, M.; Lottman,
  B.; Oomkens, P.; Rutten, N.
\newblock Plausibility of Human Remote Driving: Human-Centered Experiments from
  the Point of View of Teledrivers and Telepassengers.
\newblock {\em Proceedings of the Human Factors and Ergonomics Society Annual
  Meeting} {\bf 2019}, {\em 63},~2018–2023.
\newblock {\url{https://doi.org/10.1177/1071181319631006}}.

\bibitem[Bout et~al.(2017)Bout, Brenden, Klingeg{\aa}rd, Habibovic, and
  B{\"o}ckle]{Bou2017}
Bout, M.; Brenden, A.P.; Klingeg{\aa}rd, M.; Habibovic, A.; B{\"o}ckle, M.P.
\newblock A Head-Mounted Display to Support Teleoperations of Shared Automated
  Vehicles.
\newblock In Proceedings of the Proceedings of the 9th International Conference
  on Automotive User Interfaces and Interactive Vehicular Applications Adjunct,
  New York, NY, USA,  2017; AutomotiveUI '17, pp. 62--66.
\newblock {\url{https://doi.org/10.1145/3131726.3131758}}.

\bibitem[Georg(2024)]{Georg2024}
Georg, J.M.
\newblock Konzeption und Langzeittest der Mensch-Maschine-Schnittstelle für
  die Teleoperation von automatisierten Fahrzeugen.
\newblock PhD thesis, Technische Universität München,  2024.

\bibitem[Tener and Lanir(2022)]{Tener2022}
Tener, F.; Lanir, J.
\newblock Driving from a Distance: Challenges and Guidelines for Autonomous
  Vehicle Teleoperation Interfaces.
\newblock In Proceedings of the Proceedings of the 2022 CHI Conference on Human
  Factors in Computing Systems, New York, NY, USA,  2022; CHI '22.
\newblock {\url{https://doi.org/10.1145/3491102.3501827}}.

\bibitem[Graf and Hussmann(2020)]{Graf2020}
Graf, G.; Hussmann, H.
\newblock User Requirements for Remote Teleoperation-based Interfaces.
\newblock In Proceedings of the 12th International Conference on Automotive
  User Interfaces and Interactive Vehicular Applications, New York, NY, USA,
  2020; AutomotiveUI '20, p. 85–88.
\newblock {\url{https://doi.org/10.1145/3409251.3411730}}.

\bibitem[Gafert et~al.(2023)Gafert, Mirnig, Fr\"{o}hlich, Kraut, Anzur, and
  Tscheligi]{Gafert2023}
Gafert, M.; Mirnig, A.G.; Fr\"{o}hlich, P.; Kraut, V.; Anzur, Z.; Tscheligi, M.
\newblock Effective remote automated vehicle operation: a mixed reality
  contextual comparison study.
\newblock {\em Personal and Ubiquitous Computing} {\bf 2023}, {\em
  27},~2321–2338.
\newblock {\url{https://doi.org/10.1007/s00779-023-01782-5}}.

\bibitem[Bodell and Gulliksson(2016)]{Bodell2016}
Bodell, O.; Gulliksson, E.
\newblock Teleoperation of autonomous vehicle.
\newblock  2016.

\bibitem[Lindgren and Larsson~Vahlberg(2023)]{lindgren2023}
Lindgren, I.; Larsson~Vahlberg, A.
\newblock Remote Operation in a Modern Context {\bf 2023}.

\bibitem[Kettwich et~al.(2021)Kettwich, Schrank, and Oehl]{Kettwich2021}
Kettwich, C.; Schrank, A.; Oehl, M.
\newblock Teleoperation of Highly Automated Vehicles in Public Transport:
  User-Centered Design of a Human-Machine Interface for Remote-Operation and
  Its Expert Usability Evaluation.
\newblock {\em Multimodal Technologies and Interaction} {\bf 2021}, {\em
  5},~26.
\newblock {\url{https://doi.org/10.3390/mti5050026}}.

\bibitem[Schrank et~al.(2024)Schrank, Walocha, Brandenburg, and
  Oehl]{Schrank2024}
Schrank, A.; Walocha, F.; Brandenburg, S.; Oehl, M.
\newblock Human-centered design and evaluation of a workplace for the remote
  assistance of highly automated vehicles.
\newblock {\em Cogn. Technol. Work} {\bf 2024}, {\em 26},~183--206.

\bibitem[Tener and Lanir(2025)]{Tener2025}
Tener, F.; Lanir, J.
\newblock Guiding, not driving: Design and Evaluation of a Command-Based User
  Interface for Teleoperation of Autonomous Vehicles,  2025,
  \href{http://arxiv.org/abs/2502.00750}{{\normalfont
  [arXiv:cs.HC/2502.00750]}}.

\bibitem[Brand et~al.(2024)Brand, Baumann, and Schmitz]{Brand2024}
Brand, T.; Baumann, M.; Schmitz, M.
\newblock Bridging system limits with human--machine-cooperation.
\newblock {\em Cogn. Technol. Work} {\bf 2024}, {\em 26},~341--360.

\bibitem[Brecht et~al.(2024)Brecht, Gehrke, Kerbl, Krauss, Majstorović, Pfab,
  Wolf, and Diermeyer]{Brecht2024}
Brecht, D.; Gehrke, N.; Kerbl, T.; Krauss, N.; Majstorović, D.; Pfab, F.;
  Wolf, M.M.; Diermeyer, F.
\newblock Evaluation of Teleoperation Concepts to Solve Automated Vehicle
  Disengagements.
\newblock {\em IEEE Open Journal of Intelligent Transportation Systems} {\bf
  2024}, {\em 5},~629--641.
\newblock {\url{https://doi.org/10.1109/OJITS.2024.3468021}}.

\bibitem[Schrank et~al.(2024)Schrank, Merat, Oehl, and Wu]{Schrank2024a}
Schrank, A.; Merat, N.; Oehl, M.; Wu, Y.
\newblock Human factors considerations of remote operation supporting level 4
  automation. In {\em Lecture Notes in Mobility}; Springer Nature Switzerland,
  2024; pp. 111--125.

\bibitem[Skogsmo et~al.(2023)Skogsmo, Andersson, Jernberg, and
  Aramrattana]{Skogsmo2023}
Skogsmo, I.; Andersson, J.; Jernberg, C.; Aramrattana, M.
\newblock One2Many: remote operation of multiple vehicles,  2023.

\bibitem[Cruise(2021)]{Cruise2021}
Cruise.
\newblock Cruise Under the Hood 2021: From Self-Driving R{\&}D to Self-Driving
  Reality.
\newblock Available online:
  \url{https://youtu.be/YnhYeTWvfB0?si=1n-4RupxTqFGUSI3{\&}t=448} (accessed on
  2025-01-29).

\bibitem[Motional(2022)]{Motional2022}
Motional.
\newblock Motional's Remote Vehicle Assistance (RVA).
\newblock Available online: \url{https://www.youtube.com/watch?v=pyoHeEcgHFA}
  (accessed on 2024-01-19).

\bibitem[{Motional}(2023)]{Motional2023}
{Motional}.
\newblock Smart Choices: How Robotaxis Partner with Remote Vehicle Operators to
  Work Through Tricky Spots Safely.
\newblock Available online: \url{https://motional.com/news/rva} (accessed on
  2024-01-19).

\bibitem[Zoox(2020)]{Zoox2020}
Zoox.
\newblock How Zoox Uses TeleGuidance to Provide Remote Assistance to its
  Autonomous Vehicles.
\newblock Available online: \url{https://www.youtube.com/watch?v=NKQHuutVx78}
  (accessed on 2025-01-29).

\bibitem[Kettwich and Dre\ss{}ler(2020)]{Kettwich2020}
Kettwich, C.; Dre\ss{}ler, A.
\newblock Requirements of Future Control Centers in Public Transport.
\newblock In Proceedings of the 12th International Conference on Automotive
  User Interfaces and Interactive Vehicular Applications, New York, NY, USA,
  2020; AutomotiveUI '20, p. 69–73.
\newblock {\url{https://doi.org/10.1145/3409251.3411726}}.

\bibitem[für Normung~e. V.(2020)]{DIN9241-210}
für Normung~e. V., D.D.I.
\newblock {DIN} {EN} {ISO} 9241-210:2020-03, {Ergonomie} der
  {Mensch}-{System}-{Interaktion} - {Teil} 210: {Menschzentrierte} {Gestaltung}
  interaktiver {Systeme}; {Deutsche} {Fassung}.
\newblock Technical report, Beuth Verlag GmbH,  2020.
\newblock {\url{https://doi.org/10.31030/3104744}}.

\bibitem[Bangor et~al.(2009)Bangor, Kortum, and Miller]{Bangor2009}
Bangor, A.; Kortum, P.; Miller, J.
\newblock Determining what individual SUS scores mean: adding an adjective
  rating scale.
\newblock {\em J. Usability Studies} {\bf 2009}, {\em 4},~114–123.

\bibitem[Hinderks et~al.()Hinderks, Schrepp, and Thomaschewski]{UEQ}
Hinderks, A.; Schrepp, M.; Thomaschewski, J.
\newblock User experience questionnaire.
\newblock Available online: \url{https://www.ueq-online.org/} (accessed on
  30.01.2025).

\bibitem[Majstorović et~al.(2022)Majstorović, Hoffmann, Pfab, Schimpe, Wolf,
  and Diermeyer]{Maj2022}
Majstorović, D.; Hoffmann, S.; Pfab, F.; Schimpe, A.; Wolf, M.M.; Diermeyer,
  F.
\newblock Survey on Teleoperation Concepts for Automated Vehicles.
\newblock In Proceedings of the 2022 IEEE International Conference on Systems,
  Man, and Cybernetics (SMC),  2022, pp. 1290--1296.
\newblock {\url{https://doi.org/10.1109/SMC53654.2022.9945267}}.

\bibitem[Cohen(1992)]{Cohen1992}
Cohen, J.
\newblock A power primer.
\newblock {\em Psychological Bulletin} {\bf 1992}, {\em 112},~155–159.
\newblock {\url{https://doi.org/10.1037/0033-2909.112.1.155}}.

\end{thebibliography}

\PublishersNote{}

\end{document}